\documentstyle[aps,preprint,epsfig]{revtex}
\def\dd{{\rm d}}
\def\be{\begin{equation}}
\def\ee{\end{equation}}
\def\bea{\begin{eqnarray}}
\def\eea{\end{eqnarray}}

\begin{document}
\draft
\preprint{gr-qc/9901027}

\title{{\Large\bf Geodesic Motions in $2+1$ Dimensional \\
Charged Black Holes }}
\author{
Dong Hyun Park\thanks{\it E-mail address :
donghyun$@$newton.skku.ac.kr}
 and Seung-ho Yang\thanks{\it E-mail address : anka$@$newton.skku.ac.kr}}
\address{
Department of Physics and Institute of Basic Science,
Sungkyunkwan University\\
Suwon 440-746, Korea}
\maketitle

\begin{abstract}
We study the geodesic motions of a test particle around $2+1$ dimensional
charged black holes. We obtain a class of exact geodesic motions for the
massless test particle when the ratio of its energy and angular 
momentum is given by square root of cosmological constant. 
The other geodesic motions for both massless and massive test
particles are analyzed by use of numerical method.
\end{abstract}

\vspace{5mm}


\newpage

\section{Introduction}

Since Ba${\tilde{\rm n}}$ados-Teitelboim-Zanelli(BTZ) reported the
three dimensional black hole~\cite{btz} as a series of solutions in $2+1$ 
dimensional anti-de Sitter gravity~\cite{dj,kkk}, it has become one of the
most exciting problems in theoretical gravity.
Black hole thermodynamics and statistical properties of BTZ black holes 
have been representative topics~\cite{many,car}.
Recently the importance of BTZ-type black holes is emphasized because it
has been demonstrated that the duality between gravity in $N+1$
dimensional anti-de Sitter space and conformal field theory in $N$ 
dimensions~\cite{mal,wit}.
Among various branches of black hole researches, the simplest but basic
topic is to investigate the classical geodesic motions in $2+1$
dimensional BTZ black holes. Though the exact solutions of geodesic
motions were found for Schwarzschild- and Kerr-type BTZ 
black holes~\cite{fgscmp}, no such solutions are known for a charged 
BTZ black hole. 
These aspects seem to be similar for the other research fields, e.g.,
black hole thermodynamics~\cite{many}.
It has been believed by the following reason: The metric of a charged
BTZ black hole involves both logarithm and the square of the radial
coordinates. In this note, we found a class
of exact geodesic motions for a charged BTZ black hole despite of the
above obstacle.
In addition, all other possible geodesic motions are
categorized in examining the orbit equation, and analyzed by use of 
numerical method.

In next section, we briefly recapitulate charged BTZ black holes, and
discuss  the both null and time-like geodesics. 
We obtain a class of exact geodesic motions for the massless
test particle when the ratio of its energy and angular momentum is given
by square root of the absolute value of a negative cosmological constant.
We conclude in Sec.III with a brief discussion.

\section{Geodesic Motions}

A static $2+1$ dimensional metric with rotational symmetry has the form;
\begin{equation}
\dd s^2= B(r)e^{2N(r)}\dd t^2 -B^{-1}(r)\dd r^2 -r^2 \dd \theta^2.
\end{equation}
If there exists an electric point charge at the origin,
the electrostatic field is given by $E_r=q/r$, and the diagonal
components of energy-momentum tensor are non-vanishing, i.e.,
${T^t}_t = {T^r}_r = -{T^\theta}_\theta = {E_r}^2/2 e^{2N(r)}$.
Then the Einstein equations become
\be\label{e-e1}
\frac{1}{r} \frac{\dd N(r)}{\dd r} =  0,
\ee
\be\label{e-e2}
\frac{1}{r}\frac{\dd B(r)}{\dd r} =
 2 |\Lambda| - \frac{8 \pi G q^2}{r^2 e^{2N(r)}}.
\ee
Static solutions of Eqs. (\ref{e-e1}) and (\ref{e-e2}) are 
\begin{equation}
N(r) = N_0,
\end{equation}
\begin{equation}
B(r) = |\Lambda| r^2 - 8 \pi G q^2 \ln r - 8 G M,
\end{equation}
where we have two integration constants $N_0$ and $M$.
Note that the integration constant $N_0$ can be absorbed by rescaling
of the time variable so that one can set it to be zero. 
The other constant $M$ is identified by the mass of a BTZ black
hole\cite{BY}. 
The obtained solutions are categorized into three classes
characterized by the value of mass parameter $M$ for a given value of
charger $q$:
({\rm{\romannumeral 1}}) When $M < \left( {\pi} q^2 /{2} \right) \left[1-
\ln \left( 4 \pi G q^2 /{|\Lambda|} \right) \right]$, 
the spatial manifold does not contain a horizon.
({\rm{\romannumeral 2}}) When $M = \left( {\pi} q^2 /{2} \right) \left[1-   
\ln \left( 4 \pi G q^2 /{|\Lambda|} \right) \right]$, it has one horizon
at $r= \sqrt{ {4 \pi G q^2}/{|\Lambda|}}$ and then it corresponds to the
extremal case of a charged BTZ black hole.
({\rm{\romannumeral 3}}) When $M > \left( {\pi} q^2 /{2} \right) \left[1-
\ln \left( 4 \pi G q^2 /{|\Lambda|} \right) \right]$,
there are two horizons of a charged BTZ black hole.

Let us consider geodesic equations around the charged BTZ black hole.
There are two constants of motions, $\gamma$ and $L$, associated with two
Killing vectors such as
\begin{eqnarray}
B(r) \frac{\dd  t}{\dd s}=\gamma, \label{t-ge}\\
r^2\frac{\dd\theta}{\dd s} = L. \label{a-ge}
\end{eqnarray}
Geodesic equation for radial motions is read from the Lagrangian 
for a test particle:
\be\label{rel-lag}
B \left(\frac{\dd t}{\dd s} \right)^2 - \frac{1}{B} \left( 
\frac{\dd r}{\dd s} \right)^2 - r^2 \left( \frac{\dd \theta}{\dd s}
\right) =  m^2 \nonumber
\ee 
where $m=0$ stands for null (photon) geodesics and $m>0$ time-like
geodesics so that $m(>0)$ sets to be $1$ without loss of generality.
Inserting Eqs. (\ref{t-ge}) and (\ref{a-ge})
into Eq.~(\ref{rel-lag}), we have a first-order equation
\begin{equation}\label{r-ge}
-\frac{1}{2}\left(\frac{\dd r}{\dd s}\right)^2 =
-\frac{1}{2}\left\{B(r)\left(\frac{L^2}{r^2}+ m^2\right)
-\gamma^2\right\}.
\end{equation}
Then, all possible geodesic motions are classified by the shape of
effective potential from the right-hand side of Eq.~(\ref{r-ge}):
\begin{equation}\label{eff-pot}
V(r) = \frac{1}{2}\left\{B(r)\left(\frac{L^2}{r^2}+ m^2\right)
- {\gamma^2}\right\}.
\end{equation}
From  Eqs. (\ref{a-ge}) and (\ref{r-ge}), orbit equation is  
\begin{equation}\label{ang-mot}
\left(\frac{\dd r}{\dd \theta}\right)^2 = -B(r) r^2 \left( 1
+ \frac{m^2}{L^2} r^2 \right) +\frac{\gamma^2}{L^2}r^4.
\end{equation}

From now on let us examine the orbit equation (\ref{ang-mot}) and
analyze all possible geodesic motions for various parameters.
In the case of a photon without angular momentum ($m=0$ and $L=0$), 
the effective potential (\ref{eff-pot}) becomes a constant:
\begin{equation}
V(r)= -\frac{\gamma^2}{2}.
\end{equation}
For the regular case, all possible geodesic motions resemble those of a
free particle. 
These solutions do not depend on both electric charge $q$ and black hole
mass $M$.
For a black hole, the geodesic motions are similar to those
of regular case far away from the horizon, however the existence of black
hole horizons should be
taken into account.  
Specifically the photon also has a free particle motion near the horizon,
but the redshift is detected at the outside of black hole. 

When a test photon carries angular momentum ($m=0$ and $L\neq0$), 
the effective potential (\ref{eff-pot}) is
\begin{equation}\label{ep-photon}
V(r)= \frac{1}{2}B(r)\left(\frac{L^2}{r^2}  \right) -\frac{\gamma^2}{2},
\end{equation}
and the corresponding orbit equation is written as
\begin{equation}\label{ang-reg}
\dd \theta = \frac{\dd r}{ r \sqrt{4 \pi G q^2 \ln r^2 + 8 M G + r^2
\left( \frac{\gamma^2}{L^2} - |\Lambda| \right)}}. 
\end{equation}
There have been reported several well-known analytic
solutions of geodesic equation for Schwarzschild- or Kerr-type BTZ
black holes\cite{fgscmp} because the orbit equations include the terms
of the power of radial coordinates alone.
Once we look at the form of the orbit equation in Eq.~(\ref{ang-reg})
with both the square of the radial coordinate and logarithmic terms, 
we may easily accept non-existence 
of analytic solutions of Eq.~(\ref{ang-reg}) when the electric charge
$q$ is nonzero. 
However, a clever but simple investigation shows an exit when $\gamma / L
= \sqrt{|\Lambda|}$ in addition to the trivial Schwarzschild-type BTZ
black hole in the limit of zero electric charge ($q=0$): The coefficient
of $r^2$-term in the integrand vanishes and then a set of explicit orbits
solution seems to exist. We will show that it is indeed the case. 

As shown in FIGs.~{\ref{fig1}} and {\ref{fig2}}, all the geodesic motions
of a photon in a charged BTZ black hole are categorized by five cases:
({\rm{\romannumeral 1}}) When ${\gamma}/{L} < \left( {\gamma}/{L}
\right)_{\rm cr}$, there is no allowed motion. Every orbit is allowed only
when $\gamma / L$ is equal to or larger than the critical value
$\left( {\gamma}/{L}\right)_{\rm cr}$:
\begin{equation}\label{fac-cr}
\left(\frac{\gamma}{L}\right)_{\rm cr} = \sqrt{ |\Lambda|- \exp \left(
\frac{2 M}{\pi q^2} + \ln (4 \pi G q^2)-1 \right)}.   
\end{equation}
({\rm{\romannumeral 2}}) When ${\gamma}/{L} = \left( {\gamma}/{L}
\right)_{\rm cr}$, this condition gives a circular motion.
The radius of this circular motion is
\begin{equation}
r_{\rm cir} = \sqrt{
\frac{4 \pi G q^2}{|\Lambda|-\left(\frac{\gamma}{L}\right)^2_{\rm cr}}}.
\end{equation}
Both critical value and radius of this circular motion are obtained
from Eq.~(\ref{ep-photon}).
({\rm{\romannumeral 3}}) When $\left({\gamma}/{L}\right)_{\rm
cr}<{\gamma}/{L}< \sqrt{|\Lambda|}$, the photon has
elliptic motions, but for the charged BTZ black hole with two horizon,
the lower bound is limited by zero.
({\rm{\romannumeral 4}}) When ${\gamma}/{L}= \sqrt{|\Lambda|}$,
the geodesic equation becomes integrable. These geodesic motions are
unbounded spiral motions at a large scale.
({\rm{\romannumeral 5}}) When ${\gamma}/{L} > \sqrt{|\Lambda|}$,
the geodesic motions are unbounded.
Note that for any charged BTZ black hole, $(\gamma/L)_{\rm cr}$ in
Eq.~(\ref{fac-cr}) becomes imaginary, and then the circular orbit is
not allowed. It is also true for the extremal charged BTZ black hole.

\begin{figure}
\setlength{\unitlength}{0.1bp}
\begin{picture}(3600,2160)(0,0)
{\footnotesize
\put(2025,170){\makebox(0,0){$r$}}
\put(240,1230){\makebox(0,0)[b]{\shortstack{$V(r)$}}}
\put(3114,300){\makebox(0,0){$12$}}
\put(2243,300){\makebox(0,0){$8$}}
\put(1371,300){\makebox(0,0){$4$}}
\put(500,300){\makebox(0,0){$0$}}
\put(450,2060){\makebox(0,0)[r]{$3$}}
\put(450,1691){\makebox(0,0)[r]{$2$}}
\put(450,1322){\makebox(0,0)[r]{$1$}}
\put(450,953){\makebox(0,0)[r]{$0$}}
\put(450,584){\makebox(0,0)[r]{$-1$}}
\put(3600,1170){\makebox(0,0)[l]{$\gamma / L = 0$}}
\put(3600,1080){\makebox(0,0)[l]{$\gamma / L = (\gamma /L)_{\rm cr}$}}
\put(3600,980){\makebox(0,0)[l]
{$(\gamma / L)_{\rm cr}\!\! <\!\! \gamma /L \!\! <\!\! \sqrt{|\Lambda|}$}}
\put(3600,860){\makebox(0,0)[l]{$\gamma /L = \sqrt{|\Lambda|}$}}
\put(3600,690){\makebox(0,0)[l]{$\gamma /L > \sqrt{|\Lambda|}$}}
}
\end{picture}
\caption{\footnotesize 
The schematic shapes of effective potential $V(r)$ for various values of
$\gamma /L$ when {$M /|\Lambda| =-3$, $q^2 / |\Lambda|=1$,
 and $G=1$}. Since $M/|\Lambda| < \left(\pi q^2 / 2 \right) \left[ 1 -
\ln \left(4 \pi G q^2 /|\Lambda| \right) \right] = -3$,
the corresponding metric does not have a horizon.
}
\label{fig1}
\end{figure}
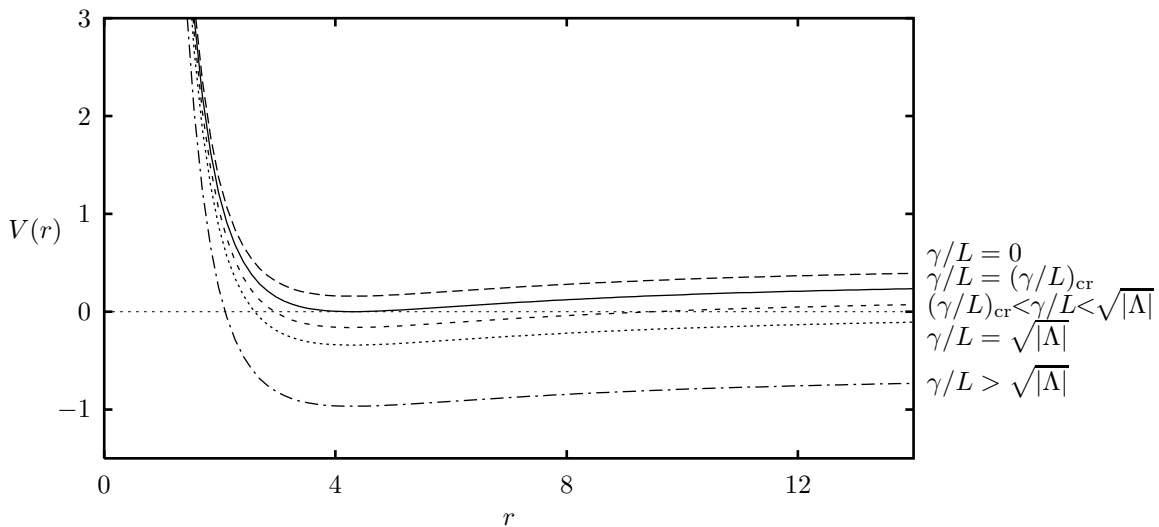
\begin{figure}
\setlength{\unitlength}{0.1bp}
\begin{picture}(3600,2160)(0,0)
{\footnotesize
\put(2025,150){\makebox(0,0){$r$}}
\put(210,1230){\makebox(0,0)[b]{\shortstack{$V(r)$}}
}
\put(3550,300){\makebox(0,0){$9$}}
\put(2533,300){\makebox(0,0){$6$}}
\put(1517,300){\makebox(0,0){$3$}}
\put(500,300){\makebox(0,0){$0$}}
\put(450,2060){\makebox(0,0)[r]{$4$}}
\put(450,1645){\makebox(0,0)[r]{$2$}}
\put(450,1230){\makebox(0,0)[r]{$0$}}
\put(450,815){\makebox(0,0)[r]{$-2$}}
\put(450,400){\makebox(0,0)[r]{$-4$}}
{
\put(3600,1340){\makebox(0,0)[l]{$\gamma/L=0$}}
\put(3600,1255){\makebox(0,0)[l]{$0<\gamma/L< \sqrt{|\Lambda|}$}}
\put(3600,1125){\makebox(0,0)[l]{$\gamma/L= \sqrt{|\Lambda|}$}}
\put(3600,1000){\makebox(0,0)[l]{$\gamma/L> \sqrt{|\Lambda|}$}}
}
\put(1000,1320){\makebox(0,0)[r]{$r^{\rm in}_H$}}
\put(2850,1320){\makebox(0,0)[r]{$r^{\rm out}_H$}}

\put(856,397){\line(0,1){1661}}
\put(2852,397){\line(0,1){1661}}
}
\end{picture}
\caption{\footnotesize 
The schematic shapes of effective potential $V(r)$ for various values of
$\gamma /L$ when {$M/|\Lambda|=0$, $q^2/|\Lambda|=1$, and $G=1$ }.
Since $M/|\Lambda| > \left(\pi q^2 / 2 \right) \left[ 1 -
\ln \left(4 \pi G q^2 /|\Lambda| \right) \right] = 0$,
the corresponding metric has two horizons.}
\label{fig2}
\end{figure}
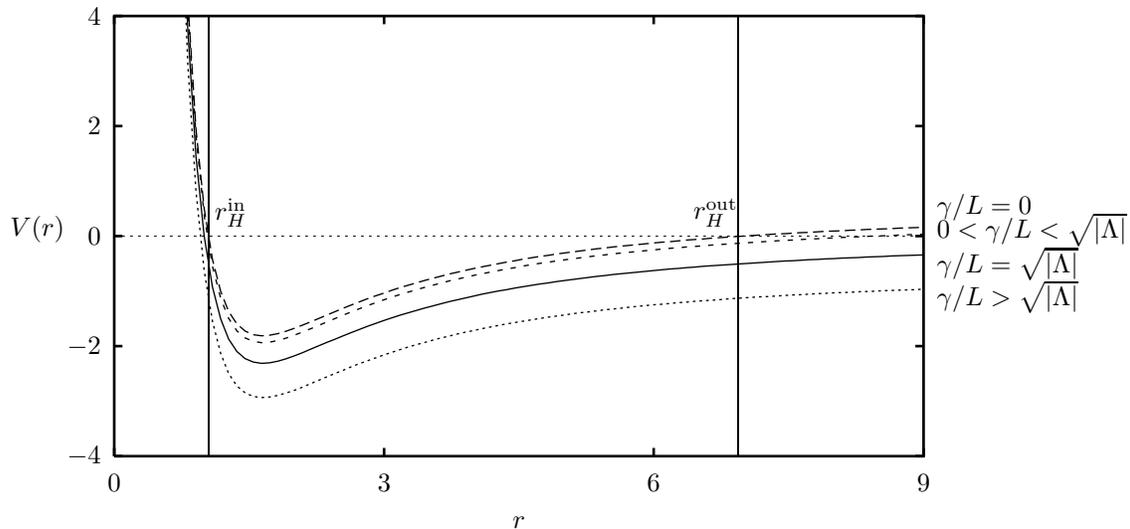
\begin{figure}
\begin{center}
\setlength{\unitlength}{0.1bp}
\begin{picture}(2123,2160)(0,0)
{\footnotesize
\put(1799,300){\makebox(0,0){$1500$}}
\put(1287,300){\makebox(0,0){$0$}}
\put(774,300){\makebox(0,0){$1500$}}
\put(450,1647){\makebox(0,0)[r]{$1500$}}
\put(450,1155){\makebox(0,0)[r]{$0$}}
\put(450,663){\makebox(0,0)[r]{$1500$}}
\put(1280,150){\makebox(0,0){(a)}}
}
\end{picture}
\setlength{\unitlength}{0.1bp}
\begin{picture}(2066,2160)(0,0)
{\footnotesize
\put(2016,300){\makebox(0,0){$300$}}
\put(1625,300){\makebox(0,0){$200$}}
\put(1233,300){\makebox(0,0){$100$}}
\put(842,300){\makebox(0,0){$0$}}
\put(450,300){\makebox(0,0){$100$}}
\put(400,1910){\makebox(0,0)[r]{$200$}}
\put(400,1532){\makebox(0,0)[r]{$100$}}
\put(400,1155){\makebox(0,0)[r]{$0$}}
\put(400,777){\makebox(0,0)[r]{$100$}}
\put(400,400){\makebox(0,0)[r]{$200$}}
\put(1280,150){\makebox(0,0){(b)}}
}
\end{picture}
\end{center}
\caption{\footnotesize 
(a). The photon trajectory falls into
the black hole at large scale, and the region bounded by a small square is
magnified in figure (b). 
(b). This figure shows incoming and outgoing photon
trajectories, and the region bounded by a small square is
displayed in FIG.~{\ref{fig4}}.
The coordinates of both figures are reduced to logarithmic scale,
when $M/|\Lambda|=-3$, $q^2/|\Lambda|=1$, $G=1$, and $\gamma / L =
\sqrt{|\Lambda|}$.
}
\label{fig3}
\end{figure}
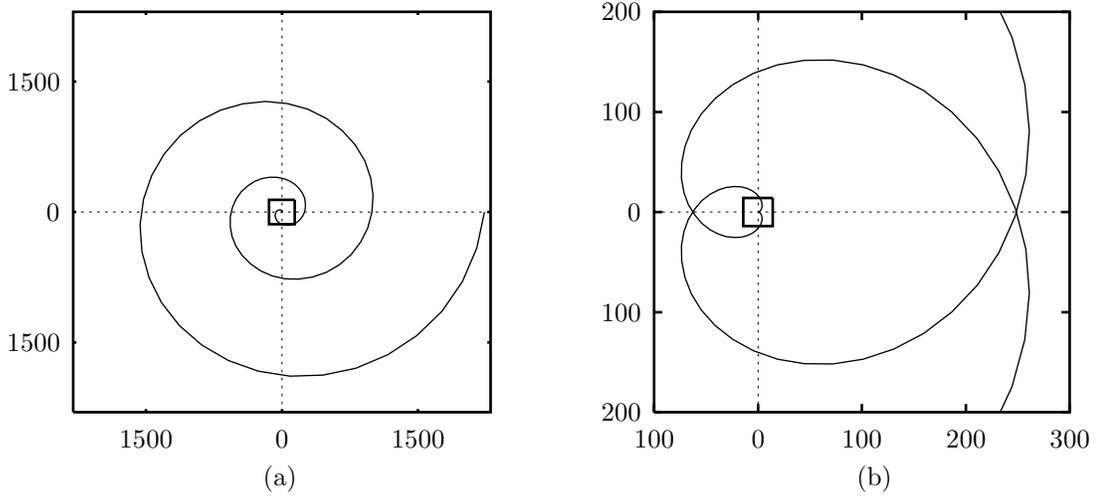
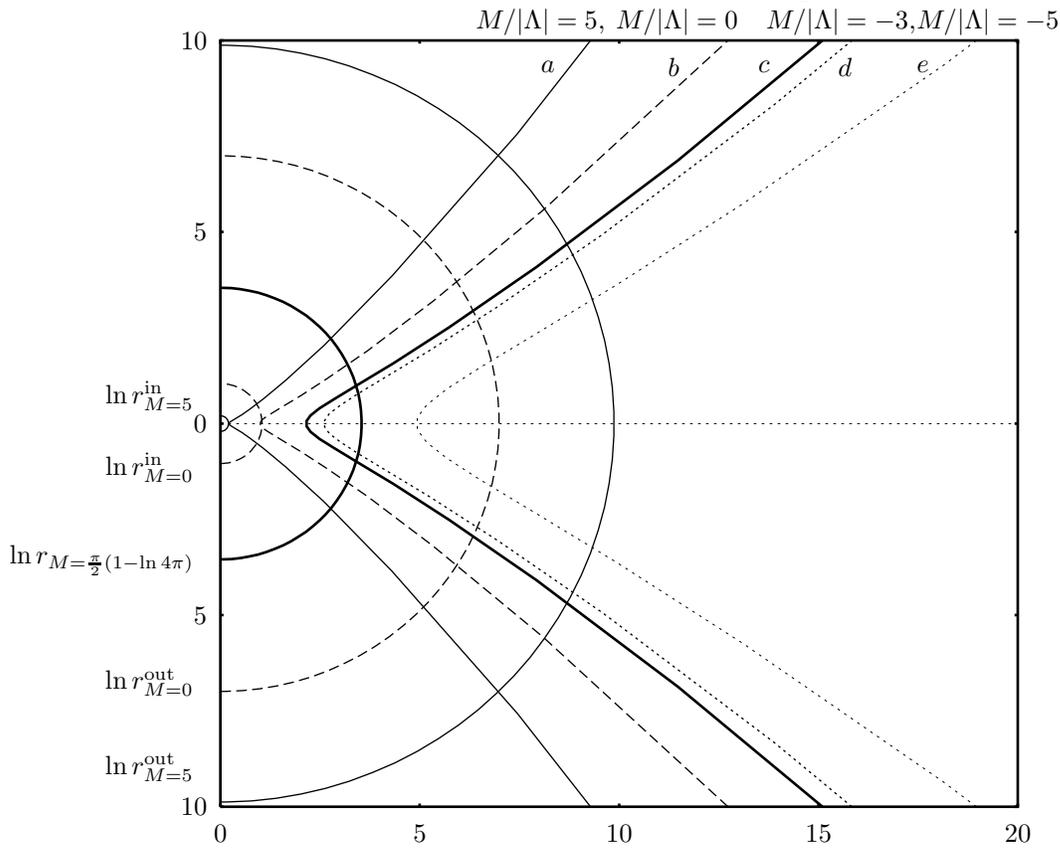
\begin{figure}
\begin{center}
\setlength{\unitlength}{0.1bp}
\begin{picture}(3455,3240)(0,0)
{\footnotesize
\put(3405,150){\makebox(0,0){$20$}}
\put(2654,150){\makebox(0,0){$15$}}
\put(1903,150){\makebox(0,0){$10$}}
\put(1151,150){\makebox(0,0){$5$}}
\put(400,150){\makebox(0,0){$0$}}
\put(350,3140){\makebox(0,0)[r]{$10$}}
\put(350,2418){\makebox(0,0)[r]{$5$}}
\put(350,1695){\makebox(0,0)[r]{$0$}}
\put(350,973){\makebox(0,0)[r]{$5$}}
\put(350,250){\makebox(0,0)[r]{$10$}}

\put(1850,3210){\makebox(0,0)[r]{$M/|\Lambda|=5$,}}
\put(2350,3210){\makebox(0,0)[r]{$M/|\Lambda|=0$}}
\put(3100,3350){\makebox(0,0)[r]{$M/|\Lambda|=\frac{\pi}{2}(1-\ln{4\pi})$}}
\put(3020,3210){\makebox(0,0)[r]{$M/|\Lambda|=-3$,}}
\put(3560,3210){\makebox(0,0)[r]{$M/|\Lambda|=-5$}}
\put(1660,3040){\makebox(0,0)[r]{$a$}}
\put(2130,3040){\makebox(0,0)[r]{$b$}}
\put(2470,3040){\makebox(0,0)[r]{$c$}}
\put(2780,3040){\makebox(0,0)[r]{$d$}}
\put(3070,3040){\makebox(0,0)[r]{$e$}}
 }
{\footnotesize
\put(300,1800){\makebox(0,0)[r]{$\ln r^{\rm in}_{M=5}$}}
\put(300,1530){\makebox(0,0)[r]{$\ln r^{\rm in}_{M=0}$}} 
\put(300,1180){\makebox(0,0)[r]{$\ln r_{M=\frac{\pi}{2}(1-\ln{4\pi})}$}} 
\put(300,720){\makebox(0,0)[r]{$\ln r^{\rm out}_{M=0}$}} 
\put(300,400){\makebox(0,0)[r]{$\ln r^{\rm out}_{M=5}$}} 
}
\end{picture}
\end{center}
\caption{\footnotesize 
This figure shows the photon trajectories when $M/|\Lambda|$ is 
changed. This figure is also reduced to logarithmic scale, when 
$q^2/|\Lambda|=1$ and $G=1$. The line (a) and (b) show the trajectories
of charged BTZ black hole, the line (c) is that of extremal case of 
charged BTZ black hole, and the line (d) and (e) indicate the regular
case. Each circles are horizons and the coordinate of this figure is also
reduced to logarithmic scale.
}
\label{fig4}
\end{figure}

We have already mentioned that Eq.~(\ref{ang-reg}) becomes integrable
when $\gamma / L = \sqrt{|\Lambda|}$.
The explicit form of the integrable orbits is
\begin{equation}\label{exact}
r = \exp \left(2 \pi G q^2 \theta^2 - \frac{M}{\pi q^2} \right),
\end{equation}
and FIGs.~{\ref{fig3}} and {\ref{fig4}} show an example.
FIG.~{\ref{fig4}} shows representative trajectories which are changed by
the mass parameter with a fixed charge, $q^2 / |\Lambda|=1$.
All possible motions are spiral at the large scale 
(see Fig.~{\ref{fig3}}-(a)). As $M/|\Lambda|$ becomes sufficiently large,
the radius of inner horizon approaches zero and that of outer horizon goes
to infinity.
In this limit, mass parameter determines the black hole dominantly and
the charge does not affect much.
Therefore, it leads to the Schwarzschild type black hole.
When the black hole mass converges to that of extremal black hole case,
i.e., $M \rightarrow ({\pi} q^2 / 2)
\left[ 1 - \ln ({4 \pi G q^2} / {|\Lambda|} ) \right]$,
the radii of inner and outer horizons are merged into one;
$$r^{\rm ext}_{H}= \sqrt{\frac{4 \pi G q^2}{|\Lambda|}}.$$
The perihelion of these analytically-obtained orbits in Eq.~(\ref{exact})
is trivially obtained
\be
r_{\rm ph} = \exp\left( - \frac{M}{\pi q^2} \right),
\ee
and, for an extremal charged BTZ black hole, it becomes
\be
r^{\rm ext}_{\rm ph}= \exp \left[\frac{1}{2}
\left( \ln \frac{4 \pi G q^2}{|\Lambda|} - 1 \right) \right].
\ee
The perihelion of these analytically-obtained orbits in Eq.~(\ref{exact})
is trivially obtained
\be
r_{\rm ph} = \exp\left( - \frac{M}{\pi q^2} \right),
\ee
and, for an extremal charged BTZ black hole, it becomes
\be
r^{\rm ext}_{\rm ph}= \exp \left[\frac{1}{2}   
\left( \ln \frac{4 \pi G q^2}{|\Lambda|} - 1 \right) \right].
\ee

\begin{figure}
\setlength{\unitlength}{0.1bp}
\begin{picture}(2066,2160)(0,0)
{\footnotesize
\put(2016,300){\makebox(0,0){$10$}}
\put(1612,300){\makebox(0,0){$5$}}
\put(1208,300){\makebox(0,0){$0$}}
\put(804,300){\makebox(0,0){$5$}}
\put(400,300){\makebox(0,0){$10$}}
\put(350,2060){\makebox(0,0)[r]{$10$}}
\put(350,1645){\makebox(0,0)[r]{$5$}}
\put(350,1230){\makebox(0,0)[r]{$0$}}
\put(350,815){\makebox(0,0)[r]{$5$}}
\put(350,400){\makebox(0,0)[r]{$10$}}
\put(1208,180){\makebox(0,0){(a)}}
}
\end{picture}
\setlength{\unitlength}{0.1bp}
\begin{picture}(2066,2160)(0,0)
{\footnotesize
\put(2016,300){\makebox(0,0){$6$}}
\put(1600,300){\makebox(0,0){$3$}}
\put(1183,300){\makebox(0,0){$0$}}
\put(767,300){\makebox(0,0){$3$}}
\put(350,300){\makebox(0,0){$6$}}
\put(300,2060){\makebox(0,0)[r]{$6$}}
\put(300,1645){\makebox(0,0)[r]{$3$}}
\put(300,1230){\makebox(0,0)[r]{$0$}}
\put(300,815){\makebox(0,0)[r]{$3$}}
\put(300,400){\makebox(0,0)[r]{$6$}}
\put(1208,180){\makebox(0,0){(b)}} 
}
\end{picture}
\caption{\footnotesize (a). The figure shows a photon trajectory of
charged BTZ black hole when $ q^2 / |\Lambda| = 1 $, $M / |\Lambda| = 0$,
$m=0$, $L = 1$, and $\gamma = 0.5 $. (b). The figure shows that of an
extremal charged BTZ black hole when $q^2 / |\Lambda| = 1$, $M / |\Lambda|
= \frac{\pi}{2} \left( 1- \ln 4 \pi \right)$, $ m=0$, $L=1$, and $\gamma =
0.5 $. The dashed circles are all horizons.}
\label{fig5}
\end{figure}
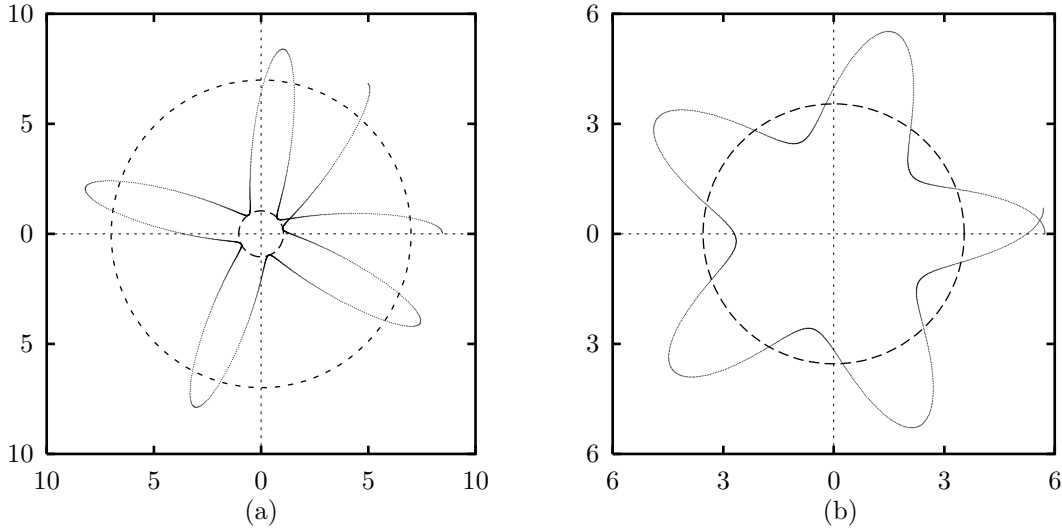

As we mentioned previously, there remain two classes of solutions:
({\rm{\romannumeral 1}}) $\left({\gamma}/{L}\right)_{\rm
cr}<{\gamma}/{L}< \sqrt{|\Lambda|}$ and
({\rm{\romannumeral 2}}) ${\gamma}/{L} > \sqrt{|\Lambda|}$.
When ${\gamma}/{L} \ne \sqrt{|\Lambda|}$, the orbit equation
(\ref{ang-reg}) is not integrable. Then the numerical analysis is a
useful tool for those geodesic motions.
For the first case ($\left({\gamma}/{L}\right)_{\rm
cr}<{\gamma}/{L}< \sqrt{|\Lambda|}$), all orbits are bounded between 
aphelion and perihelion. Two representative examples of elliptic geodesic
motions are shown in FIG.~{\ref{fig5}}.
For the second case (${\gamma}/{L} > \sqrt{|\Lambda|}$), FIGs.~\ref{fig1}
and \ref{fig2} show that there also exists perihelion but we
cannot obtain analytically. 

For the motions of a massive particle ($m=1$), all allowed motions
are bounded since the asymptotic structure of spacetime is not flat,
but anti-de Sitter. 

In the case of a massive test particle with zero angular 
momentum ($m = 1$ and $L = 0$), the effective potential becomes 
\begin{equation}
V(r) = \frac{1}{2}B(r) - \frac{\gamma^2}{2}.
\end{equation}
There is no allowed motion under the critical energy $\gamma_{\rm cr}$.
When the minimum value of the effective potential is zero, the 
critical energy of the test particle is computed;
\begin{equation}
\gamma_{\rm cr}= \sqrt{4 \pi G q^2 \left(1-\ln\frac{4 \pi G
q^2}{|\Lambda|} \right) - 8 G M}.
\end{equation}
When $\gamma=\gamma_{\rm cr}$, the test particle remains at rest.
Above the critical energy, radial motion is an oscillation between 
perihelion and aphelion.
For the black hole case, the motion of a test particle is also
oscillating, but its range is restricted by the horizons.

\begin{figure}
\setlength{\unitlength}{0.1bp}
\begin{picture}(3600,2160)(0,0)
{\footnotesize
\put(2050,150){\makebox(0,0){$r$}}
\put(100,1230){\makebox(0,0)[b]{\shortstack{$V(r)$}}}

\put(2900,660){\makebox(0,0)[r]{$r_{\rm ap}$}}
\put(1250,660){\makebox(0,0)[r]{$r_{\rm ph}$}}
\put(2050,660){\makebox(0,0)[r]{$r_{\rm cir}$}}

}
{\footnotesize
\put(3550,300){\makebox(0,0){$8$}}
\put(2800,300){\makebox(0,0){$6$}}
\put(2050,300){\makebox(0,0){$4$}}
\put(1300,300){\makebox(0,0){$2$}}
\put(550,300){\makebox(0,0){$0$}}
\put(500,2060){\makebox(0,0)[r]{$40$}}
\put(500,1728){\makebox(0,0)[r]{$30$}}
\put(500,1396){\makebox(0,0)[r]{$20$}}
\put(500,1064){\makebox(0,0)[r]{$10$}}
\put(500,732){\makebox(0,0)[r]{$0$}}
\put(500,400){\makebox(0,0)[r]{$-10$}}


\put(3850,1370){\makebox(0,0)[r]{$\gamma = 0$}}
\put(3925,1265){\makebox(0,0)[r]{$\gamma = \gamma_{\rm cr}$}}
\put(3925,1070){\makebox(0,0)[r]{$\gamma > \gamma_{\rm cr}$}}

}
\end{picture}
\caption{\footnotesize  
The schematic shapes of effective potential $V(r)$ for various values of
$\gamma$ when {$M/|\Lambda|=-3$,
$q^2/|\Lambda|=1$; $L=1$, and $m=1$}. Since $M/|\Lambda|=-3$, the
corresponding metric does not have a horizon.
}
\label{fig6}
\end{figure}
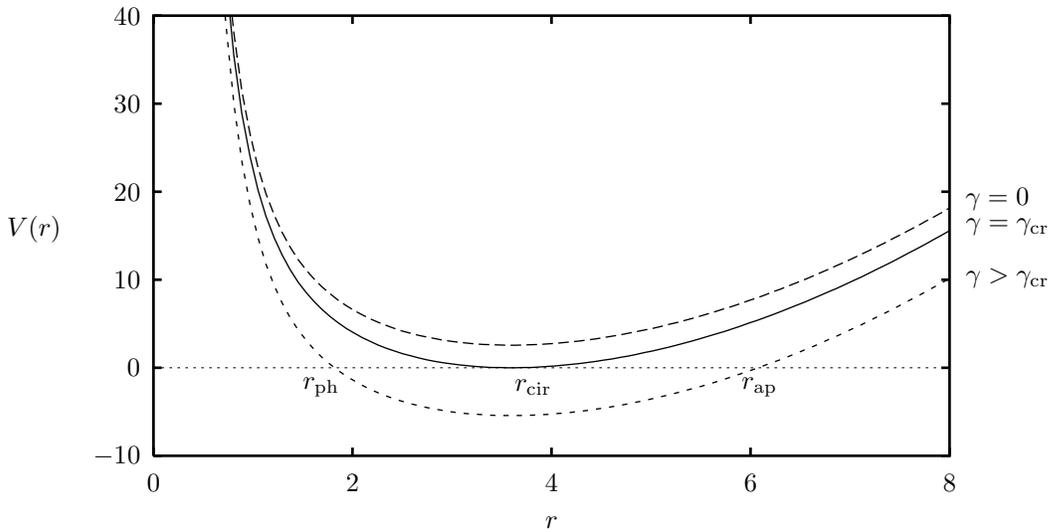

\begin{figure}
\setlength{\unitlength}{0.1bp}
\begin{picture}(3600,2160)(0,0)
{\footnotesize
\put(2050,150){\makebox(0,0){$r$}}
\put(100,1230){\makebox(0,0)[b]{\shortstack{$V(r)$}}}
}
{\footnotesize
\put(3550,300){\makebox(0,0){$10$}}
\put(2950,300){\makebox(0,0){$8$}}
\put(2350,300){\makebox(0,0){$6$}}
\put(1750,300){\makebox(0,0){$4$}}
\put(1150,300){\makebox(0,0){$2$}}
\put(550,300){\makebox(0,0){$0$}}
\put(500,2060){\makebox(0,0)[r]{$30$}}
\put(500,1728){\makebox(0,0)[r]{$20$}}
\put(500,1396){\makebox(0,0)[r]{$10$}}
\put(500,1064){\makebox(0,0)[r]{$0$}}
\put(500,732){\makebox(0,0)[r]{$-10$}}
\put(500,400){\makebox(0,0)[r]{$-20$}}
}
{\footnotesize
\put(3900,1810){\makebox(0,0)[r]{$\gamma=0$}}
\put(3900,1640){\makebox(0,0)[r]{$\gamma>0$}}
}
{\footnotesize
\put(1000,1140){\makebox(0,0)[r]{$r^{\rm in}_H$}}
\put(2645,1140){\makebox(0,0)[r]{$r^{\rm out}_H$}}
}
\put(863,397){\line(0,1){1661}}
\put(2650,397){\line(0,1){1661}}
\end{picture}
\caption{\footnotesize 
The schematic shapes of effective potential $V(r)$ for various values of
$\gamma$ when {$M/|\Lambda|=0$, $q^2/|\Lambda|=1$, $L=1$, and $m=1$}.
Since $M/|\Lambda| = 0$, the corresponding metric has two horizons. 
}
\label{fig7}
\end{figure}
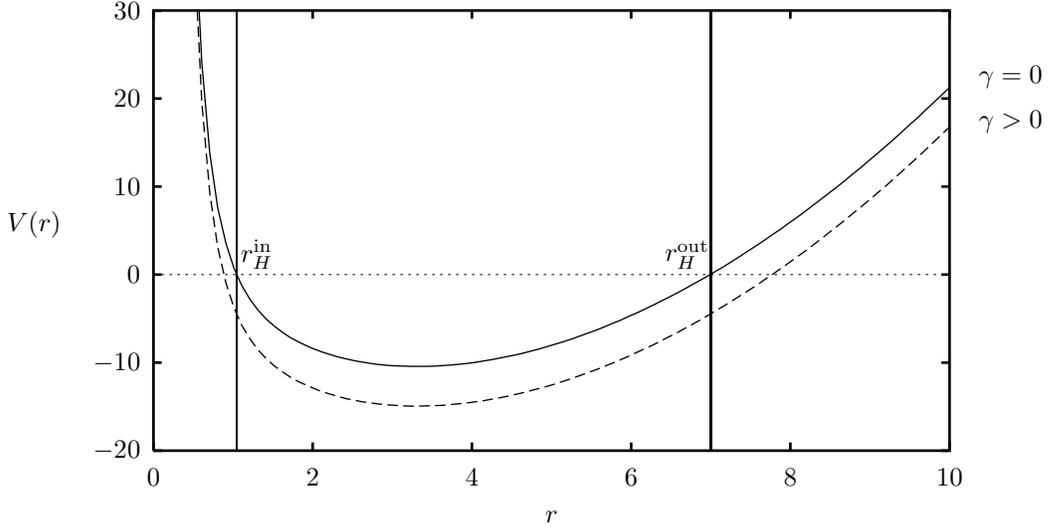

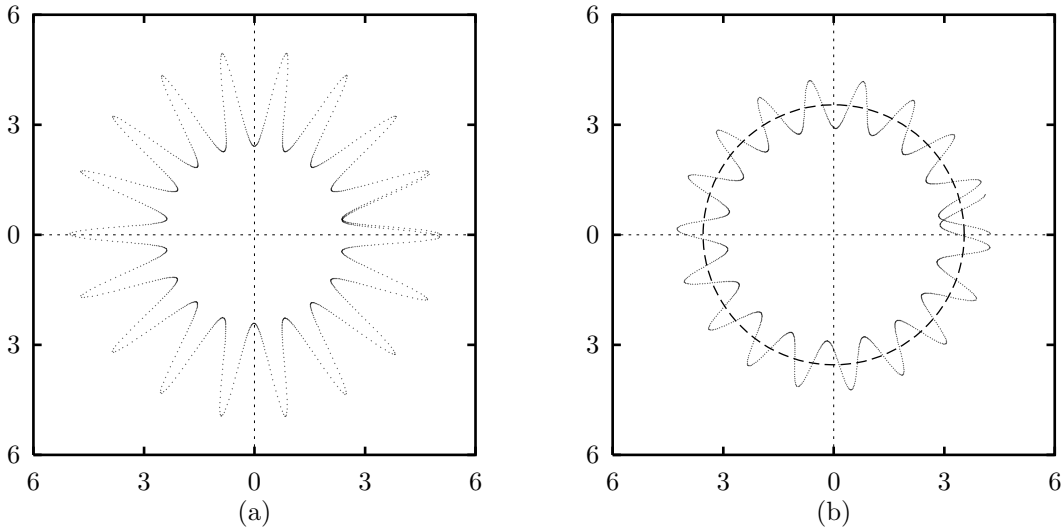
\begin{figure}
\setlength{\unitlength}{0.1bp}
\begin{picture}(2066,2160)(0,0)
{\footnotesize
\put(2016,300){\makebox(0,0){$6$}}
\put(1600,300){\makebox(0,0){$3$}}
\put(1183,300){\makebox(0,0){$0$}}
\put(767,300){\makebox(0,0){$3$}}
\put(350,300){\makebox(0,0){$6$}}
\put(300,2060){\makebox(0,0)[r]{$6$}}
\put(300,1645){\makebox(0,0)[r]{$3$}}
\put(300,1230){\makebox(0,0)[r]{$0$}}
\put(300,815){\makebox(0,0)[r]{$3$}}
\put(300,400){\makebox(0,0)[r]{$6$}}
\put(1183,180){\makebox(0,0){(a)}}
}
\end{picture}
\setlength{\unitlength}{0.1bp}
\begin{picture}(2066,2160)(0,0)
{\footnotesize
\put(2016,300){\makebox(0,0){$6$}}
\put(1600,300){\makebox(0,0){$3$}}
\put(1183,300){\makebox(0,0){$0$}}
\put(767,300){\makebox(0,0){$3$}}
\put(350,300){\makebox(0,0){$6$}}
\put(300,2060){\makebox(0,0)[r]{$6$}}
\put(300,1645){\makebox(0,0)[r]{$3$}}
\put(300,1230){\makebox(0,0)[r]{$0$}}
\put(300,815){\makebox(0,0)[r]{$3$}}
\put(300,400){\makebox(0,0)[r]{$6$}}
\put(1183,180){\makebox(0,0){(b)}}
}
\end{picture}
\vspace{-0.3cm}
\caption{\footnotesize (a). The figure shows a elliptic motion 
of regular case when $M/|\Lambda|=-3$ and $\gamma=3$. (b).
The figure shows that of extremal charged BTZ black hole when
$M/|\Lambda| = ({\pi} / {2}) \left[ 1- \ln ( {4 \pi G} / {|\Lambda|})
\right]$ and $\gamma=1$. 
The other values are same for $q^2/|\Lambda|=1$, $L=1$, $m=1$, and
$G=1$. The dashed circle in figure (b) indicates a extremal horizon.}
\label{fig8}
\end{figure}

In the case of a massive test particle with angular 
momentum ($m = 1$ and $L \neq 0$), the effective potential becomes
\begin{equation}
V(r) = \frac{1}{2} B(r) \left( \frac{L^2}{r^2} + 1  \right)
- \frac{\gamma^2}{2}.
\end{equation}
FIG.~{\ref{fig6}} and FIG.~{\ref{fig7}} depict effective potentials for 
various values of $\gamma$:
FIG.~{\ref{fig6}} corresponds to a regular spacetime and 
FIG.~{\ref{fig7}} a charged BTZ black hole.  
For the regular case, the dashed line shows that the minimum of $V(r)$
is positive and then there is no allowed motion below this critical
energy $\gamma_{\rm cr}$. It is the same as that of the $L=0$ case.
When $\gamma = \gamma_{\rm cr}$, the minimum of $V(r)$ is zero and then
there exists a circular motion at $r = r_{\rm cir}$ (see the solid line in
FIG.~{\ref{fig6}}).   
The effective potential given by the dotted line in FIG.~{\ref{fig6}}    
supports the elliptic motion with aphelion $r_{\rm ap}$ and perihelion
$r_{\rm ph}$. The effective potentials for a charged BTZ black hole are
shown in FIG.~{\ref{fig7}}.
As shown in FIG.~{\ref{fig7}}, the motions outside the horizons are
provided only when $\gamma > 0$.
The unique allowed motion for the extremal charged BTZ black hole is
the stopped motion at the degenerated horizon, which means eventually
that no motion is allowed.
Two examples of the trajectories of a massive test particle are described
in FIG.~{\ref{fig8}}.
\vspace{-0.2cm}

\section{conclusion}

In this paper we have studied the geodesic motions of charged BTZ
black holes. We found a class of exact geodesic solutions
of a massless test particle when the ratio of its energy and angular
momentum is equal to the square root of the absolute value of a negative
cosmological constant.
The obtained geodesics describe the unbounded spiral motion.
Though we have some exact geodesic motions, it seems impossible for us
to extend our coordinates to Kruskal-Szekeres or Penrose diagram
which provide a basis for further researches.
We categorized the possible geodesic motions of massive and massless
test particles as circular, elliptic, unbounded spiral, and unbounded 
motions. Several typical examples are analyzed by numerical works.
Many works in various field, e.g., black hole thermodynamics,
have been done for Schwarzschild- or Kerr-type BTZ black
holes~\cite{many,car,kal}. On the other hand, those researches have been 
limited in the case of charged BTZ black holes, that is different from
that of $3+1$ dimensional Reissner-Nordstr${\rm \ddot{o}}$m black holes.
We hope that our simple work provides a building block to further
researches about charged BTZ black holes and related topics.

\acknowledgments{
The authors would like to thank Yoonbai Kim for helpful discussions.
This work was supported by KRF(1998-015-D00075) and KOSEF through Center
for Theoretical Physics, SNU.}

\end{document}